\documentclass[twocolumn, aps, prb]{revtex4}
\usepackage{graphicx}
\usepackage{amssymb}
\usepackage{amsmath}
\def\Journal #1,#2,#3,#4#5#6#7{#1 {\bf #2}, #3 (#4#5#6#7)}

\def\gsim{\lower -0.3ex \hbox{$>$} \kern -0.75em \lower 0.7ex
\hbox{$\sim$}}
\def\lsim{\lower -0.3ex \hbox{$<$} \kern -0.75em \lower 0.7ex
\hbox{$\sim$}}

\def\Vec#1{{\bf #1}}
\def\GVec#1{\mbox{\boldmath $#1$}}

\begin{document}
%%%%%%%%%%%%%%%%%%%%%%%%%%%%%%%%%%%%%%%%%%%%%%%%%%%%%%%%%%%%%%%%%%%%%%%%%%%%%%%
%
%%%%%%%%%%%%%%%%%%%%%%%%%%%%%%%%%%%%%%%%%%%%%%%%%%%%%%%%%%%%%%%%%%%%%%%%%%%%%%%
\title{
Optical Absorption in Twisted Bilayer Graphene
}
\author{Pilkyung Moon and Mikito Koshino}
\affiliation{
Department of Physics, Tohoku University, 
Sendai, 980--8578, Japan}
\date{\today}

\begin{abstract}
We theoretically study the optical absorption property
of twisted bilayer graphenes with various stacking geometries,
and demonstrate that the spectroscopic characteristics serve as a fingerprint 
to identify the rotation angle between two layers.
We find that the absorption spectrum almost continuously evolves
in changing the rotation angle, 
regardless of the lattice commensurability.
The spectrum is characterized by series of peaks 
associated with the van Hove singularity,
and the peak energies systematically shift
with the rotation angle.
We calculate the optical absorption in two different frameworks;
the tight-binding model and the effective continuum model
based on the Dirac equation.
For small rotation angles less than $10^\circ$, 
the effective model well reproduces 
the low-energy band structure and the optical conductivity 
of the tight-binding model, 
and also explains the optical selection rule analytically 
in terms of the symmetry of the effective Hamiltonian.

\end{abstract}
\maketitle

%%%%%%%%%%%%%%%%%%%%%%%%%%%%%%%%%%%%%%%%%%%%%%%%%%%%%%%%%%%%%%%%%%%%%%%%%%%%%%%
%
%%%%%%%%%%%%%%%%%%%%%%%%%%%%%%%%%%%%%%%%%%%%%%%%%%%%%%%%%%%%%%%%%%%%%%%%%%%%%%%
\section{INTRODUCTION}
\label{sec_introduction}

\begin{figure}
\begin{center}
\leavevmode\includegraphics[width=0.9\hsize]{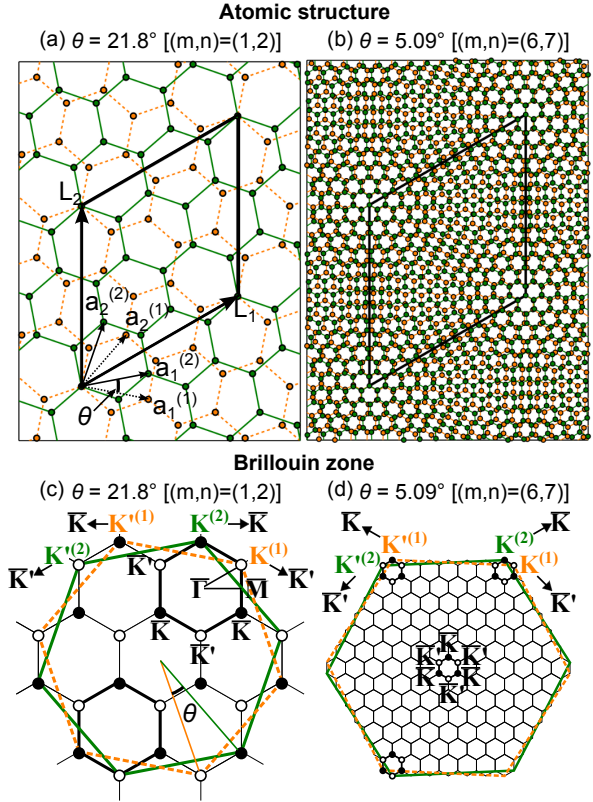}
\end{center}
\caption{Atomic structures of TBGs with
(a) $\theta= 21.8^\circ$ and (b) $\theta=5.09^\circ$.
Dashed (orange) and solid (green) lines represent
the lattices of layers 1 and 2, respectively. 
Brillouin zone of TBGs with
(c) $\theta= 21.8^\circ$ and (d) $\theta=5.09^\circ$.
Dashed (orange) and solid (green) large hexagons 
indicate the first Brillouin zone of layer 1 and 2, respectively, 
and thick small-hexagon is the folded Brillouin zone of TBG. 
}
\label{fig_atomic_structure_and_brillouin_zone}
\end{figure}

The recent advances in fabrication of atomically-thin materials
\cite{berger2006electronic,ni2008reduction,yan2011growth,xie2011graphene,zhao2011scanning}
have realized a new kind of two-dimensional superlattice
in which the lattice mismatch between neighboring layers
gives rise to an additional potential modulation.
One example of such systems is 
twisted bilayer graphene (TBG),
in which two graphene layers are stacked
with an arbitrary orientation.\cite{berger2006electronic,hass2007structural,hass2008multilayer,li2009observation,miller2010structural,luican2011single}
In TBG, the interlayer interaction between two misoriented layers
significantly modifies the low-energy band structure,
arising novel electronic features distinct from intrinsic graphene.
In decreasing the rotation angle, 
the interference between two lattice periods
produces a Moir\'{e} pattern with a long wavelength, 
where the characteristic features such as band gaps
and van Hove singularity appear in the far-infrared region,
and the band velocity of Dirac cone is significantly reduced.
\cite{lopes2007graphene,hass2008multilayer,ni2008reduction,morell2010flat,shallcross2010electronic,trambly2010localization,bistritzer2011moirepnas,PhysRevB.86.155449}
Recently, the band properties of TBG has been 
probed by Raman spectroscopy,
\cite{ni2008reduction,ni2009g,righi2011graphene,sato2012zone}
optical reflection spectroscopy,\cite{wang2010stacking}
angle-resolved photoemission spectroscopy,\cite{PhysRevB.85.075415}
and by terahertz time-domain spectroscopy.\cite{PhysRevLett.110.067401}

\begin{figure*}
%\begin{figure}
\begin{center}
%\leavevmode\includegraphics[width=\hsize]{fig_commensurate_and_incommensurate_atomic_structures.eps}
\leavevmode\includegraphics[width=0.8\hsize]{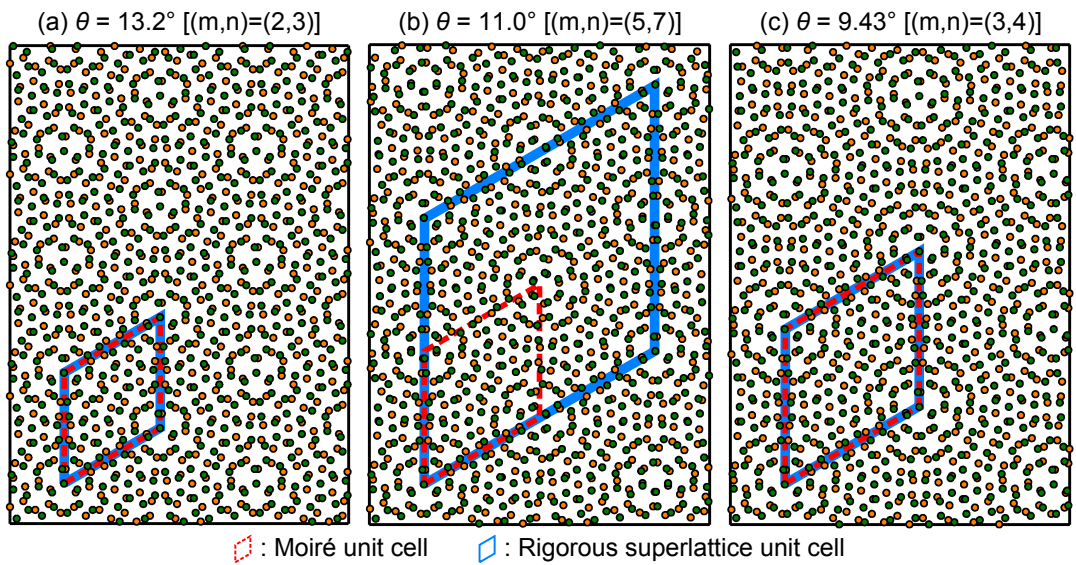}
\end{center}
\caption{
Atomic structures of TBGs with (a) $\theta=13.2^\circ$,
(b) $11.0^\circ$, and (c) $9.43^\circ$.
Dashed (red) and solid (blue) parallelograms correspond to
the Moir\'{e} unit cell and rigorous superlattice unit cell, respectively.}
\label{fig_commensurate_and_incommensurate_atomic_structures}
%\end{figure}
\end{figure*}

The purpose of this paper is to reveal the optical 
absorption properties of TBGs with various stacking geometries.
The optical absorption measurement is widely adopted
for graphene-based systems
to investigate the electronic structures.
\cite{nair2008fine,li2008dirac,mak2008measurement,zhang2008determination,henriksen2008cyclotron,li2009band,
zhang2009direct,orlita2009graphite,kuzmenko2009infrared,kuzmenko2009determination,mak2010evolution,mak2010electronic}
Theoretically, the optical absorption 
for light incident perpendicular to the layer
is related to the dynamical conductivity,
which was calculated for monolayer graphene,
\cite{ando2002dynamical,gusynin2006transport,gusynin2006unusual,gusynin2007anomalous}
$AB$-stacked graphene bilayer and multilayers,
\cite{abergel2007optical,falkovsky2007optical,koshino2008magneto,koshino2009electronic,peres2010colloquium,koshino2013stacking}
and also for TBG with a specific rotation angle.
\cite{wang2010stacking, chen2011stacking}
In this paper, we calculate the dynamical conductivity
of TBG in a wide range of rotation angles, and demonstrate that 
the spectroscopic characteristics serve as a fingerprint 
to identify the stacking geometry.
We find that the spectrum is characterized by 
series of peaks associated with the van Hove singularity,
of which transition energy 
continuously shifts as the rotation angle is changed.
Here the dynamical conductivity is calculated in two different frameworks;
the tight-binding model and the effective continuum model
based on the Dirac equation.
For the latter, we develop a general treatment
to derive the effective model from arbitrary tight-binding parametrization.
For small rotation angles less than $10^\circ$, 
the effective model nicely reproduces 
the low-energy band structure and the dynamical conductivity 
of the tight-binding model, 
and also explains the optical selection rule analytically
in terms of the symmetry of the effective Hamiltonian.

\section{Theoretical methods}

\subsection{Atomic structure and Brillouin zone}
\label{subsection:Atomic structure and Brillouin zone}

Graphene is a single layer of carbon atoms
arranged in a honeycomb lattice structure,
of which unit cell includes two 
inequivalent sublattice sites, $A$ and $B$.
The stacking geometry of bilayer graphene
is characterized by the relative rotation angle $\theta$ combined with
the lateral translation $\GVec{\delta}$ between the layers.
Here we define the structure of TBG 
by rotating the layer 1 and 2 
of $AA$-stacked bilayer 
around a common $B$-site by $-\theta/2$ and $+\theta/2$, respectively,
and then translating the layer 2 relatively to the layer 1 by $\GVec{\delta}$.
We define  $\Vec{a}_1 = a(1,0)$ and $\Vec{a}_2 = a(1/2,\sqrt{3}/2)$ 
as the lattice vectors of the $AA$-stacked bilayer
before the rotation, 
where $a \approx 0.246\,\mathrm{nm}$ is the lattice constant.
The lattice vectors of the layer $l$ after the rotation
are given by $\Vec{a}_i^{(l)} =R(\mp \theta/2)\Vec{a}_i$ 
with $\mp$ for $l=1,2$, 
respectively, where $R(\theta)$ represents the rotation by $\theta$.

When $\GVec{\delta}$ is fixed to 0, 
the rotation $\theta=0$ and $60^\circ$ 
give $AA$ and $AB$ stacking, respectively.
$60^\circ - \theta$ is equivalent to
$-\theta$ followed by
a relative translation of the layer 2 from $A$ site to $B$ site.
\cite{mele2010commensuration,mele2012interlayer}
Also, $\theta$ and $-\theta$ are mirror images 
sharing equivalent band structures.
Therefore, it is reasonable to characterize the geometry of TBG
by the combination of $\theta$ ($0 \le \theta \le 30^\circ$)
and $\GVec{\delta}$.

The lattice structure of TBG is not periodic
in general angles because the periods of two graphene layers
are generally incommensurate with each layer. 
But in some special angles where two periods happen to match, 
the structure becomes rigorously periodic giving a finite unit cell.
This takes place when $\theta$ coincides with the
angle between $\Vec{v}_1 = m \Vec{a}_1 + n \Vec{a}_2$ and
$\Vec{v}_2 = n \Vec{a}_1 + m \Vec{a}_2$ with certain integers $m$ and $n$,
because then the lattice points $\Vec{v}_1$ on the layer 1
and $\Vec{v}_2$ on the layer 2 
of the non-rotated bilayer graphene 
merge after the rotation $\theta/2$ and $-\theta/2$, respectively.
The lattice vectors of the superlattice unit cell are thus given by 
\cite{mele2010commensuration}
\begin{eqnarray}
\Vec{L}_1 = m \Vec{a}_1^{(1)} + n \Vec{a}_2^{(1)}
= n \Vec{a}_1^{(2)} + m \Vec{a}_2^{(2)},
\label{eq_L_1}
\end{eqnarray}
and $\Vec{L}_2 = R(\pi/3) \Vec{L}_1$.
The rotation angle $\theta$ is related to $(m,n)$ by
\begin{eqnarray}
 \cos\theta = \frac{1}{2}\frac{m^2+n^2+4mn}{m^2+n^2+mn}.
\end{eqnarray}
The lattice constant $L = |\Vec{L}_1| = |\Vec{L}_2|$ is 
\begin{eqnarray}
 L = a \sqrt{m^2 + n^2 + mn}
= \frac{|m-n|a}{2\sin(\theta/2)}.
\label{eq_superlattice_period}
\end{eqnarray}

Figures \ref{fig_atomic_structure_and_brillouin_zone}(a) and
\ref{fig_atomic_structure_and_brillouin_zone}(b) show
the atomic structures of TBG with two different rotation angles
$\theta = 21.8^\circ$ $[(m,n)=(1,2)]$ and
$5.09^\circ$ $[(m,n)=(6,7)]$, respectively.
Unless otherwise noted, $\GVec{\delta}=0$ is used.
Figures \ref{fig_atomic_structure_and_brillouin_zone}(c)
and \ref{fig_atomic_structure_and_brillouin_zone}(d)
show the corresponding Brillouin zone in the extended scheme.
In each figure,
dashed (orange) and solid (green) large hexagons correspond to
the first Brillouin zones of layer 1 and 2, respectively,
and thick small hexagon to the folded Brillouin zone of TBG.
$K^{(l)}$ and $K'^{(l)}$ denote the two inequivalent 
valleys of layer $l$.
%, which are Dirac points in the single-layer band structure.
The four valleys $K^{(1)}$, $K'^{(1)}$, $K^{(2)}$, and $K'^{(2)}$
of the two layers are folded back to the two Dirac points,
$\bar{K}$ and $\bar{K}'$, 
in the folded Brillouin zone.\cite{shallcross2010electronic}

% Moire
When the rotation angle is small,
the mismatch between the lattice vectors of the two layers
gives rise to a Moir\'{e} pattern 
with a long  spatial period
as seen in Fig.\ \ref{fig_atomic_structure_and_brillouin_zone}(b).
\cite{shallcross2010electronic, green1948theorie}
The local lattice structure near a certain point $\Vec{r}$
approximates a non-rotated bilayer graphene 
with displacement $\GVec{\delta}$, which depends on the position as
\begin{eqnarray}
 \GVec{\delta}(\Vec{r}) = 
2\sin(\theta/2)(\Vec{e}_z \times \Vec{r}),
\label{eq_delta_of_r}
\end{eqnarray}
where $\Vec{r}$ is measured from the center of the rotation,
and $\Vec{e}_z$ is a unit vector perpendicular to the plane.
The period of the Moir\'{e} pattern $\Vec{L}^{\rm M}_i$
can be obtained by the condition that $\GVec{\delta}(\Vec{L}^{\rm M}_i)$
coincides with a primitive lattice vector of
the original $AA$-stacked bilayer. We may choose
\begin{eqnarray}
 \Vec{L}^{\rm M}_1 = \frac{(-\Vec{a}_1 +\Vec{a}_2) \times \Vec{e}_z}{2\sin(\theta/2)},
\quad
\Vec{L}^{\rm M}_2 = \frac{-\Vec{a}_1 \times \Vec{e}_z}{2\sin(\theta/2)},
\label{eq_moire_lattice_vectors}
\end{eqnarray}
giving $\GVec{\delta}(\Vec{L}^{\rm M}_1) = -\Vec{a}_1 +\Vec{a}_2$
and $\GVec{\delta}(\Vec{L}^{\rm M}_2) = -\Vec{a}_1$.
The lattice constant 
$L_{\rm M} = |\Vec{L}^{\rm M}_1|=|\Vec{L}^{\rm M}_2|$ is
\begin{eqnarray}
L_{\rm M} = \frac{a}{2 \sin(\theta/2)}.
\label{eq_moire_period}
\end{eqnarray}

The Moir\'{e} superlattice vectors $\Vec{L}^{\rm M}_i$ can be
always defined for any $\theta$, even when
the lattice structure is incommensurate.
At commensurate angles, the rigorous superlattice period $L$ 
is exactly $|m-n|$ times bigger than
the Moir\'{e} period $L_{\rm M}$.
In Fig.\ \ref{fig_commensurate_and_incommensurate_atomic_structures}, 
we illustrate the lattice structures of 
$\theta = 13.2^\circ$ $[(m,n)=(2,3)]$, $\theta = 11.0^\circ$ $[(m,n)=(5,7)]$
and
$\theta = 9.43^\circ$ $[(m,n)=(3,4)]$.
We can see that the atomic structure exactly matches the Moir\'{e} pattern
in 13.2$^\circ$ and 9.43$^\circ (|m-n|=1)$,
while in 11.0$^\circ$, 
the exact period $L$ is twice as large as $L_{\rm M}$ since $|m-n| =2$,
and accordingly the atomic structure is slightly different
between neighboring units in the Moir\'{e} pattern.

%For an example, we show the atomic structure
%of TBG at rotation angle $\theta = 21.8^\circ$ $[(m,n)=(1,2)]$
%in Figure \ref{fig_atomic_structure_and_brillouin_zone}(a),
%which has the smallest unit cell among TBG's other than $\theta=0$.

%%
%%%%%%%%%%%%%%%%%%%%%%%%%%%%%%%%%%%%%%%%

%\section{RESULTS AND DISCUSSION}
%\label{section:results_and_discussion}

\subsection{Tight-binding model}
\label{subsection:tight-binding_model}

We calculate the eigenenergies and eigenfunctions in TBG
using the tight-binding model for $p_z$ atomic orbitals.
The Hamiltonian is written as
\begin{eqnarray}
 H = -\sum_{\langle i,j\rangle}
t(\Vec{R}_i - \Vec{R}_j)
|\Vec{R}_i\rangle\langle\Vec{R}_j| + {\rm H.c.},
\label{eq_Hamiltonian_TBG}
\end{eqnarray}
where $\Vec{R}_i$ and $|\Vec{R}_i\rangle$ 
represent the lattice point and the atomic state at site $i$, respectively,
and $t(\Vec{R}_i - \Vec{R}_j)$ is
the transfer integral between the sites $i$ and $j$. 
We adopt an approximation,
\cite{nakanishi2001conductance,uryu2004electronic,trambly2010localization,slater1954simplified}
\begin{eqnarray}
 && -t(\Vec{d}) = 
V_{pp\pi}\left[1-\left(\frac{\Vec{d}\cdot\Vec{e}_z}{d}\right)^2\right]
+ V_{pp\sigma}\left(\frac{\Vec{d}\cdot\Vec{e}_z}{d}\right)^2,
\nonumber \\
&& V_{pp\pi} =  V_{pp\pi}^0 % -\tilde\gamma_0
\exp \left(- \frac{d-a_0}{\delta_0}\right),
\nonumber \\
&& V_{pp\sigma} =  V_{pp\sigma}^0 %\tilde\gamma_1
 \exp \left(- \frac{d-d_0}{\delta_0}\right),
\label{eq_transfer_integral}
\end{eqnarray}
where
%$\Vec{e}_z$ is the unit vector
%perpendicular to the graphene plane,
$a_0 = a/\sqrt{3} \approx 0.142$nm is the distance of
neighboring $A$ and $B$ sites on monolayer,
and $d_0 \approx 0.335\,\mathrm{nm}$
is the interlayer spacing.
$V_{pp\pi}^0$ is the transfer integral between 
the nearest-neighbor atoms of monolayer graphene
and $V_{pp\sigma}^0$ is that
between vertically located atoms on the neighboring layers. 
Here we take $V_{pp\pi}^0 \approx -2.7\,\mathrm{eV}$,
$ V_{pp\sigma}^0 \approx 0.48\,\mathrm{eV}$, to
fit the dispersions of monolayer graphene and $AB$-stacked bilayer 
graphene.\cite{trambly2010localization}
$\delta_0$ is the decay length of the transfer integral,
and is chosen as $0.184 a$ so that 
the next nearest intralayer coupling becomes $0.1 V_{pp\pi}^0$.
\cite{uryu2004electronic,trambly2010localization} 
The transfer integral for $d > 4 a_0$ is exponentially small 
and can be safely neglected.
%The band velocity of the Dirac cone in monolayer graphene is given by
%\begin{eqnarray}
% v \approx \frac{\sqrt{3}}{2}\frac{V_{pp\pi}^0}{\hbar}.
%\end{eqnarray}

\subsection{Effective continuum model}
\label{subsection:effective_continuum_model}

When the rotation angle is small
and the Moir\'{e} superlattice period 
is much larger than the lattice constant,
the interaction between the two graphene layers
has only the long-wavelength components,
allowing one to treat the problem in the 
effective continuum model.
The continuum approaches for TBG have been introduced 
in several literatures. \cite{lopes2007graphene,bistritzer2011moirepnas,kindermann2011local,PhysRevB.86.155449}
Here we develop a general treatment 
to construct an effective model
directly from 
the tight-binding Hamiltonian in Eq.\ (\ref{eq_Hamiltonian_TBG}).
To construct the Hamiltonian matrix, 
we define the Bloch wave basis of a single layer as
\begin{eqnarray}
&& |\Vec{k},A_l\rangle = 
\frac{1}{\sqrt{N}}\sum_{\Vec{R}_{A_l}} e^{i\Vec{k}\cdot\Vec{R}_{A_l}}
|\Vec{R}_{A_l}\rangle,
\nonumber \\
&& |\Vec{k},B_l\rangle = 
\frac{1}{\sqrt{N}}\sum_{\Vec{R}_{B_l}} e^{i\Vec{k}\cdot\Vec{R}_{B_l}}
|\Vec{R}_{B_l}\rangle,
\label{eq_bloch_base}
\end{eqnarray}
where 
the position $\Vec{R}_{A_l}(\Vec{R}_{B_l})$
runs over all $A(B)$ sites on the layer $l(= 1, 2)$,
$N$ is the number of monolayer's unit cell in the whole system,
and $\Vec{k}$ is two-dimensional Bloch wave vector 
defined in the first Brillouin zone of monolayer on the layer $l$.

Intralayer matrix element of each layer
occurs only within the same wave vector, and it is given by
\begin{eqnarray}
&& h_{A_lA_l}(\Vec{k}) 
\equiv \langle \Vec{k},A_l| H |\Vec{k},A_l\rangle
= h(\Vec{k},0),
\nonumber\\
&& h_{A_lB_l}(\Vec{k})
\equiv \langle \Vec{k},A_l| H |\Vec{k},B_l\rangle
= h(\Vec{k},\GVec{\tau}_1),
\nonumber\\
&& h_{B_lB_l}(\Vec{k})  = h_{A_lA_l}(\Vec{k}),
\label{eq_intralayer}
\end{eqnarray}
where $\GVec{\tau}_1 = (2\Vec{a}_2-\Vec{a}_1)/3$
is a vector connecting from B site to A site,
and
\begin{eqnarray}
&& h(\Vec{k},\GVec{\tau}) = 
\sum_{n_1,n_2}
- t(n_1 \Vec{a}_1 + n_2 \Vec{a}_2 + \GVec{\tau})
\nonumber\\
&&
\hspace{20mm}
\times
\exp\left[i\Vec{k}\cdot(n_1 \Vec{a}_1 + n_2 \Vec{a}_2 +  \GVec{\tau})
\right].
\end{eqnarray}
The low-energy spectrum of the monolayer graphene
is approximated by 
effective Dirac cones centered at $K$ and $K'$ points.
\cite{mcclure1956diamagnetism,divincenzo1984self,semenoff1984condensed,shon1998quantum,ando2005theory}
We take $\Vec{K} = (2\pi/a) (-2/3,0)$ and 
$\Vec{K}' = (2\pi/a) (2/3,0)$ as the $K$-points of 
non-rotated graphene.
The $K$ points of the layer $l$ are then given by 
$\Vec{K}^{(l)} = R(\mp\theta/2)\Vec{K}$ 
and $\Vec{K}'^{(l)}= R(\mp\theta/2)\Vec{K}'$, 
with $\mp$ for $l=1$ and 2, respectively.
When $\Vec{k}$ is close to either of $K$ or $K'$
the intralayer matrix element is approximately written
as \cite{ando2005theory},
\begin{eqnarray}
&& h_{A_lB_l}(\Vec{k})
\nonumber\\
&& 
\approx 
\begin{cases}
-\hbar v 
[( k_x - K^{(l)}_x )
- i ( k_y - K^{(l)}_y )
]e^{-i\eta^{(l)}} & (\Vec{k} \approx \Vec{K}),
\\
-\hbar v 
[-( k_x - K'^{(l)}_x )
- i ( k_y - K'^{(l)}_y )
]e^{i\eta^{(l)}} & (\Vec{k} \approx \Vec{K}'),
\end{cases}
\nonumber\\
\label{eq_effective_intralayer}
\end{eqnarray}
where $\eta^{(l)} = \pm\theta/2$ for $l=1$ and 2, respectively.
In the following we neglect the phase factor $e^{-i\eta^{(l)}}$
assuming $\theta \ll 1$. 
The parameter $v$ is the band velocity of the Dirac cone,
which is given in the present tight-binding parametrization as
\begin{eqnarray}
 v \approx \frac{\sqrt{3}}{2}\frac{a}{\hbar}V_{pp\pi}^0(1-2e^{-a_0/\delta_0}),
\end{eqnarray}
where the first and the second terms in the bracket 
originate from the hopping
between the first and the second nearest $AB$ pairs, respectively.
The diagonal matrix element $h_{A_lA_l}(\Vec{k}) = h_{B_lB_l}(\Vec{k})$
are shown to be of the order of $|\Vec{k}-\Vec{K}|^2$
and $|\Vec{k}-\Vec{K}'|^2$ near $K$ and $K'$ points, respectively,
and will be neglected in the following.

For the interlayer coupling,
we first consider a non-rotated bilayer graphene with $\theta = 0$
and a fixed lattice displacement $\GVec{\delta}$.
%as illustrated in Fig.\ \ref{fig_atomic_structure_translation_dependence}.
The unit cell is spanned by  monolayer's 
lattice vectors, $\Vec{a}_1$ and $\Vec{a}_2$,
which are now shared by both layers. 
As the system has the same periodicity as the monolayer,
the interlayer coupling occurs within states belonging to the same $\Vec{k}$.
The interlayer matrix element is written as
\begin{eqnarray}
&& U_{A_2A_1}(\Vec{k},\GVec{\delta}) 
\equiv \langle \Vec{k},A_2| H |\Vec{k},A_1\rangle
= u(\Vec{k},\GVec{\delta}),
\nonumber\\
&& U_{B_2B_1}(\Vec{k},\GVec{\delta}) 
\equiv \langle \Vec{k},B_2| H |\Vec{k},B_1\rangle
= u(\Vec{k},\GVec{\delta}),
\nonumber\\
&& U_{B_2A_1}(\Vec{k},\GVec{\delta}) 
\equiv \langle \Vec{k},B_2| H |\Vec{k},A_1\rangle
= u(\Vec{k},\GVec{\delta} - \GVec{\tau}_1),
\nonumber\\
&& U_{A_2B_1}(\Vec{k},\GVec{\delta}) 
\equiv \langle \Vec{k},A_2| H |\Vec{k},B_1\rangle
= u(\Vec{k},\GVec{\delta} + \GVec{\tau}_1),
\label{eq_interlayer_U}
\end{eqnarray}
where
\begin{eqnarray}
 u(\Vec{k},\GVec{\delta}) = 
\sum_{n_1,n_2}
- t(n_1 \Vec{a}_1 + n_2 \Vec{a}_2 + d_0\Vec{e}_z + \GVec{\delta})
\nonumber\\
\hspace{20mm}
\times
\exp\left[-i\Vec{k}\cdot(n_1 \Vec{a}_1 + n_2 \Vec{a}_2 + \GVec{\delta})
\right].
\end{eqnarray}
$u(\Vec{k},\GVec{\delta})$ can be immediately calculated by 
taking a summation over some small $n_i$'s, 
since $t(\Vec{d})$ rapidly vanishes in $|\Vec{d}|\gg a$.

When $\theta$ is slightly shifted from zero to a small finite angle,
the local lattice structure is approximately viewed as
a non-rotated bilayer graphene, where
the displacement $\GVec{\delta}$ slowly depends on the position
$\Vec{r}$ in accordance with Eq.\ (\ref{eq_delta_of_r}).
Then the interlayer interaction couples wave vectors
$\Vec{k}$ and $ \Vec{k}'$ which are close to each other
such that $|\Vec{k}'-\Vec{k}| \ll 2\pi/a$.
The interlayer matrix element 
is approximately written as
\begin{eqnarray}
&&\langle \Vec{k}',X'_2| H |\Vec{k},X_1\rangle
\approx\nonumber\\
&&\hspace{0mm}
\frac{1}{\Omega_{\rm M}} 
\int_{\Omega_{\rm M}} d\Vec{r}\,
U_{X'_2X_1}
\left[\frac{\Vec{k}+\Vec{k}'}{2},\GVec{\delta}(\Vec{r})
\right]
e^{-i(\Vec{k}'-\Vec{k})\cdot\Vec{r}},
\label{eq_effective_interlayer}
\end{eqnarray}
where $X$ and $X'$ are either of $A$ or $B$, 
$U_{X'_2,X_1}$ are the interlayer coupling in non-rotational
bilayer in Eq.\ (\ref{eq_interlayer_U}),
and  $\Omega_{\rm M} = |\Vec{L}^{\rm M}_1 \times \Vec{L}^{\rm M}_2|$ is the
Moir\'{e} superlattice unit cell.
The derivation of Eq.\ (\ref{eq_effective_interlayer})
is detailed in
%Appendix at the end of this section.
Appendix \ref{sec_app1}.
$U_{X'_2X_1}\left[\Vec{q},\GVec{\delta}(\Vec{r})\right]$
is periodic in $\Vec{r}$ with the Moir\'{e} superlattice periods,
and therefore the matrix element Eq.\ (\ref{eq_effective_interlayer})
is nonzero only when
$\Vec{k}'-\Vec{k} = n_1 \Vec{G}^{\rm M}_1 + n_2 \Vec{G}^{\rm M}_2$
where $\Vec{G}^{\rm M}_i$ is the reciprocal lattice vector
satisfying $\Vec{L}^{\rm M}_i \cdot \Vec{G}^{\rm M}_j =
2\pi\delta_{ij}$,
and $n_i$ is an integer.

%Using Eq.\ (\ref{eq_delta_of_r}), this is further transformed into 
%an integral in $\GVec{\delta}$ as
%\begin{eqnarray}
%&&\langle \Vec{k}',X'_2| H |\Vec{k},X_1\rangle
%\approx\nonumber\\
%&&\hspace{3mm}
%\frac{1}{\Omega'} 
%\int_{\Omega'} d\GVec{\delta}\,
%U_{X'_2X_1}
%\left[\frac{\Vec{k}+\Vec{k}'}{2},\GVec{\delta}
%\right]
%e^{-i[\Vec{e}_z\times(\Vec{k}'-\Vec{k})]
%\cdot \mbox{\scriptsize \boldmath $\delta$}/\theta}
%\nonumber\\
%\label{eq_effective_interlayer2}
%\end{eqnarray}
%where $\Omega'$ is the unit cell spanned by
%$\GVec{\delta}(\Vec{L}^M_1) = -\Vec{a}_1 +\Vec{a}_2$
%and $\GVec{\delta}(\Vec{L}^M_2) = -\Vec{a}_1$,
%and we used $2\sin(\theta/2)\approx \theta$.

In TBG, the low-energy physics is still dominated by
the states near $K$ and $K'$ points because the interlayer coupling
is much smaller than intralayer coupling.
Besides, the states near $K$ and those near $K'$
are far apart in the wave space when $\theta$ is small, 
so that they not hybridized by the interlayer coupling. 
Therefore, we may consider 
two valleys separately in constructing the Hamiltonian,
and the factor $(\Vec{k}+\Vec{k}')/2$
in Eq.\ (\ref{eq_effective_interlayer})
can be replaced with $\Vec{K}$ or $\Vec{K}'$.
In the real space representation, the effective Hamiltonian 
near $K$ is concisely written in the basis of 
$\{|A_1\rangle,|B_1\rangle,|A_2\rangle,|B_2\rangle \}$
as
\begin{eqnarray}
 H_{\rm eff} = 
\begin{pmatrix}
H_1 & U^\dagger \\
U & H_2
\end{pmatrix},
\label{eq_eff_model}
\end{eqnarray}
with 
\begin{eqnarray}
&& H_l = -\hbar v (\hat{\Vec{k}}-\Delta\Vec{K}^{(l)})\cdot \GVec{\sigma},
\nonumber\\
&& U =
\begin{pmatrix}
u(\Vec{K},\GVec{\delta})
& u(\Vec{K},\GVec{\delta}-\GVec{\tau}_1)
\\
 u(\Vec{K},\GVec{\delta}+\GVec{\tau}_1)
& u(\Vec{K},\GVec{\delta})
\end{pmatrix},
\label{eq_submatrices}
\end{eqnarray}
where $ \hat{\Vec{k}} = -i \partial/\partial\Vec{r}$,
$\GVec{\sigma} = (\sigma_x, \sigma_y)$ is the Pauli matrices,
$\Delta\Vec{K}^{(l)} = \Vec{K}^{(l)} - \Vec{K}$,
and $\GVec{\delta} = \GVec{\delta}(\Vec{r})$ is defined 
in Eq.\ (\ref{eq_delta_of_r}).
When deriving $H_l$ in Eq.\ (\ref{eq_submatrices})
from Eq.\ (\ref{eq_effective_intralayer}),
we replace $\Vec{k}$ with $\hat{\Vec{k}}+\Vec{K}$,
i.e., measure the wavenumber relatively to
the common point $\Vec{K}$ for both layers.
The Hamiltonian 
for $K'$ is obtained by replacing $\Vec{K}$ with $\Vec{K}'$
and $\hat{k}_x$ with $-\hat{k}_x$ above.
In the present choice of the tight-binding parameters
$V_{pp\pi}^0$, $V_{pp\sigma}^0$ and $\delta_0$,
the effective interlayer coupling  $u(\Vec{K},\GVec{\delta}(\Vec{r}))$ 
is approximately written 
in terms of only a few Fourier components as 
\begin{eqnarray}
 u(\Vec{K},\GVec{\delta}(\Vec{r}))  \approx 
(0.103{\rm eV}) \times \left[
1 + e^{-i \Vec{G}^{\rm M}_2\cdot \Vec{r}}
+ e^{-i (\Vec{G}^{\rm M}_1 + \Vec{G}^{\rm M}_2)\cdot \Vec{r}}
\right].
\end{eqnarray}
$u(\Vec{K}',\GVec{\delta})$ is given by $u(\Vec{K},\GVec{\delta})^*$.  
The expression of $u(\Vec{K},\GVec{\delta})$ explicitly
depends on the choice of $\Vec{K}$ vector 
out of three equivalent corners in the Brillouin zone.

In the $k$-space representation,
the Hamiltonian matrix can be written
in the space of the single-layer bases at discrete $k$-points 
$\Vec{k} = \Vec{k}_0 + n_1 \Vec{G}^{\rm M}_1 + n_2 \Vec{G}^{\rm M}_2$,
where $\Vec{k}_0$ is a vector defined in the superlattice Brillouin zone
spanned by $\Vec{G}^{\rm M}_1$ and $\Vec{G}^{\rm M}_2$.
$\Vec{k}_0 = 0$ corresponds to the $\bar{M}$ point.
To obtain the energy spectrum and eigen wave function,
we choose $k$-points satisfying $\hbar v| \Vec{k}| \lsim E_{\rm max}$ 
with a sufficiently large $E_{\rm max}$,
and diagonalize the Hamiltonian within the limited wave space.
To avoid a discrete change in the number of bases in varying $\Vec{k}_0$,
we adopt a soft cutoff which gradually reduces 
the matrix elements for the single-layer bases beyond
$E_{\rm max}$.

It is straightforward to show the Hamiltonian 
Eq.\ (\ref{eq_eff_model}) has a certain symmetry 
expressed as
\begin{eqnarray}
 \hat{\Sigma}^{-1}  H_{\rm eff} \hat{\Sigma} =  -H_{\rm eff}^*,
\label{eq_e-h_symmetry}
\end{eqnarray}
where
\begin{eqnarray}
\hat{\Sigma} =
\begin{pmatrix}
0 & \sigma_x \\
-\sigma_x & 0
\end{pmatrix}.
\end{eqnarray}
This immediately concludes that 
if $\psi$ is an eigenstate of $H_{\rm eff}$ belonging to 
energy $E$, $\hat{\Sigma} \psi^*$ is an eigenstate of energy $-E$.

\subsection{Dynamical conductivity}
\label{subsection:optical_conductivity}

Using the eigen wave functions obtained by the tight-binding model
or the effective continuum model,
we calculate the dynamical conductivity
\begin{eqnarray}
\sigma_{xx}(\omega) &=& 
\frac{e^2\hbar}{i S}
\sum_{\alpha,\beta}
\frac{f(\varepsilon_\alpha)-f(\varepsilon_\beta)}
{\varepsilon_\alpha-\varepsilon_\beta}
\frac{|\langle\alpha|v_x|\beta\rangle|^2}{\varepsilon_\alpha-\varepsilon_\beta+\hbar\omega+i\eta},
\label{eq_dynamical_conductivity}
\end{eqnarray}
where the sum is over all states,
$S$ is the area of the system, 
$f(\varepsilon)$ is the Fermi distribution function,
$\varepsilon_{\alpha}$ ($\varepsilon_{\beta}$) and
$|\alpha\rangle$ ($|\beta\rangle$)
represent the eigenenergy and the eigenstate of the system,
$v_x=-(i/\hbar)[x,H]$ is the velocity operator, and 
$\eta$ is the phenomenological broadening
which is set to $3\,\rm meV$
in the following calculations.
The optical absorption intensity at photon energies $\hbar\omega$
is related to the real part of $\sigma(\omega)$.
The transmission of light incident perpendicular to two-dimensional
system is
given by\cite{ando1975theory}
\begin{equation}
 T = \Big| 1+ \frac{2\pi}{c}\sigma_{xx}(\omega) \Big|^{-2}
\approx 1- \frac{4\pi}{c}{\rm Re}\,\sigma_{xx}(\omega).
\label{eq_Transmission}
\end{equation}
%In the following, 
%we obtain the eigenstates at $300\times300$ $k$-points
%in the folded Brillouin zone,
%then calculate the dynamical conductivity at 
%the charge neutral point
%with a phenomenological broadening of $\eta=3\,\rm meV$.

\section{Results and discussions}
%\section{Theoretical methods}
%\label{section:theoretical_methods}

\begin{figure}
\begin{center}
\leavevmode\includegraphics[width=0.9\hsize]{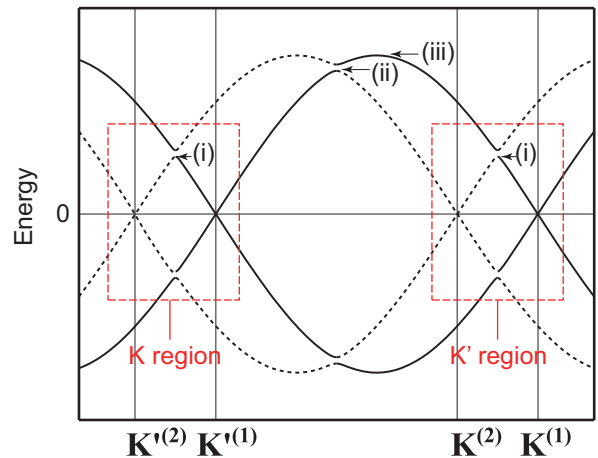}
\end{center}
\caption{Schematic band structures of TBG
in the extended zone scheme
along the line $K'^{(2)}$-$K'^{(1)}$-$K^{(2)}$-$K^{(1)}$.
Symbols (i), (ii) and (iii) indicate the saddle points.
}
\label{fig_schematic_valley_coupling}
\end{figure}

\begin{figure}
\begin{center}
\leavevmode\includegraphics[width=0.9\hsize]{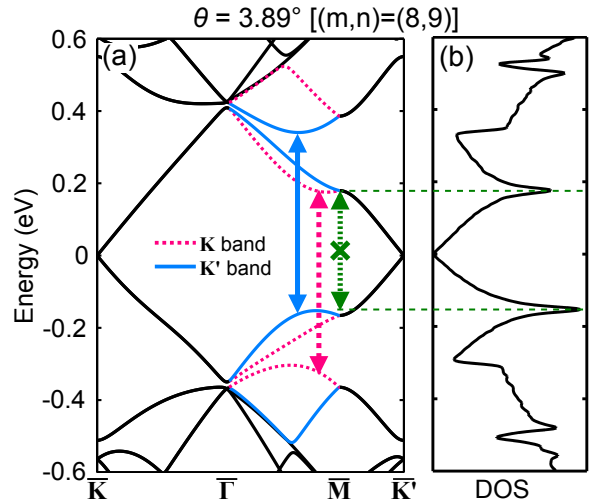}
\end{center}
\caption{(a) Band structure and (b) density of states
of TBG with $\theta= 3.89^\circ$.
Solid (blue) and dashed (pink) arrows
represent the excitation corresponding to the major peaks
in the optical absorption,
and green dotted arrow is a process optically forbidden 
(see text).}
\label{fig_band_structure_and_DOS_389}
\end{figure}

\subsection{Band structure}
\label{subsection:band_structure}

\begin{figure*}[ht]
\begin{center}
%\leavevmode\includegraphics[width=\hsize]{fig_band_structure_angle_dependence.eps}
\leavevmode\includegraphics[width=\hsize]{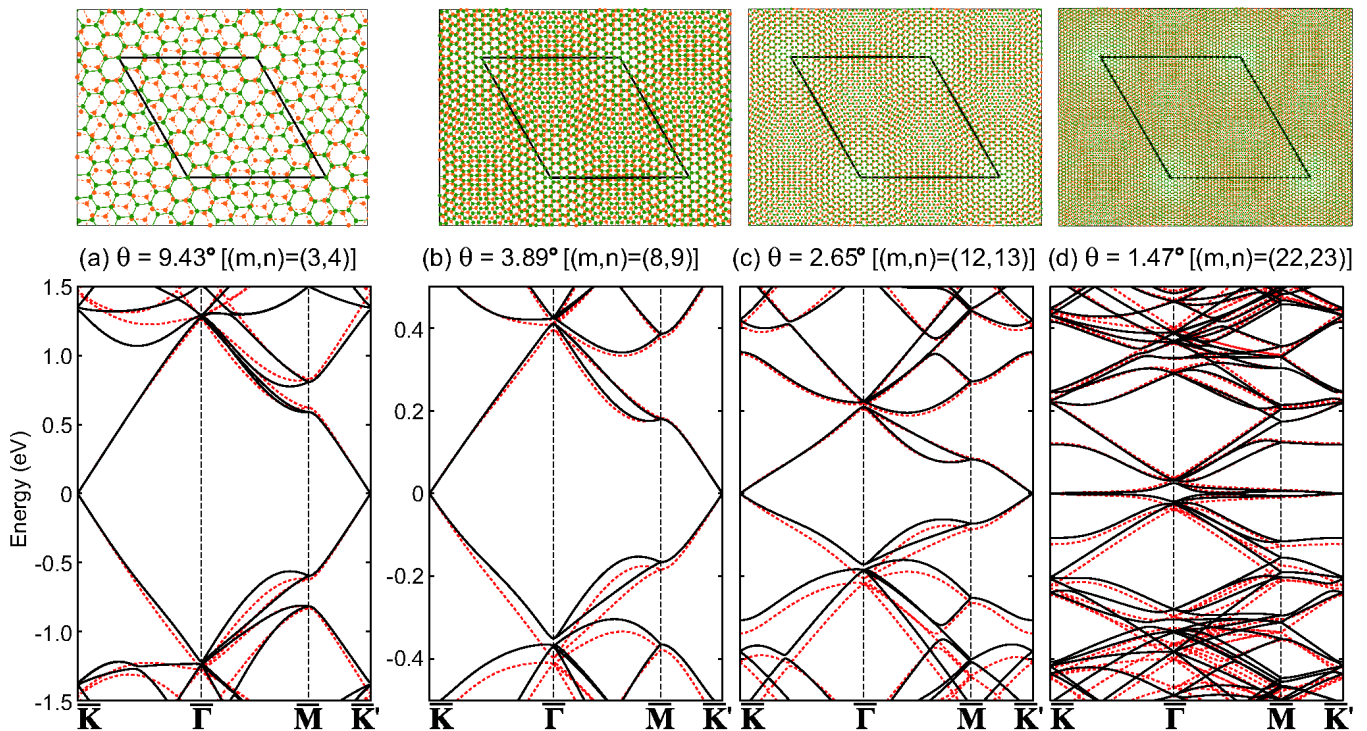}
\end{center}
\caption{Atomic structures (top) and
band structures (bottom) of TBGs with 
(a) $\theta = 9.43^\circ$, 
(b) $3.89^\circ$, (c) $2.65^\circ$, and (d) $1.47^\circ$,
calculated by the tight-binding model (solid black lines)
and the effective continuum model (dashed red lines).
Dirac point energy is set to zero.}
\label{fig_band_structure_angle_dependence}
\end{figure*}

\begin{figure*}[ht]
\begin{center}
%\leavevmode\includegraphics[width=\hsize]{fig_DOS_angle_dependence.eps}
\leavevmode\includegraphics[width=\hsize]{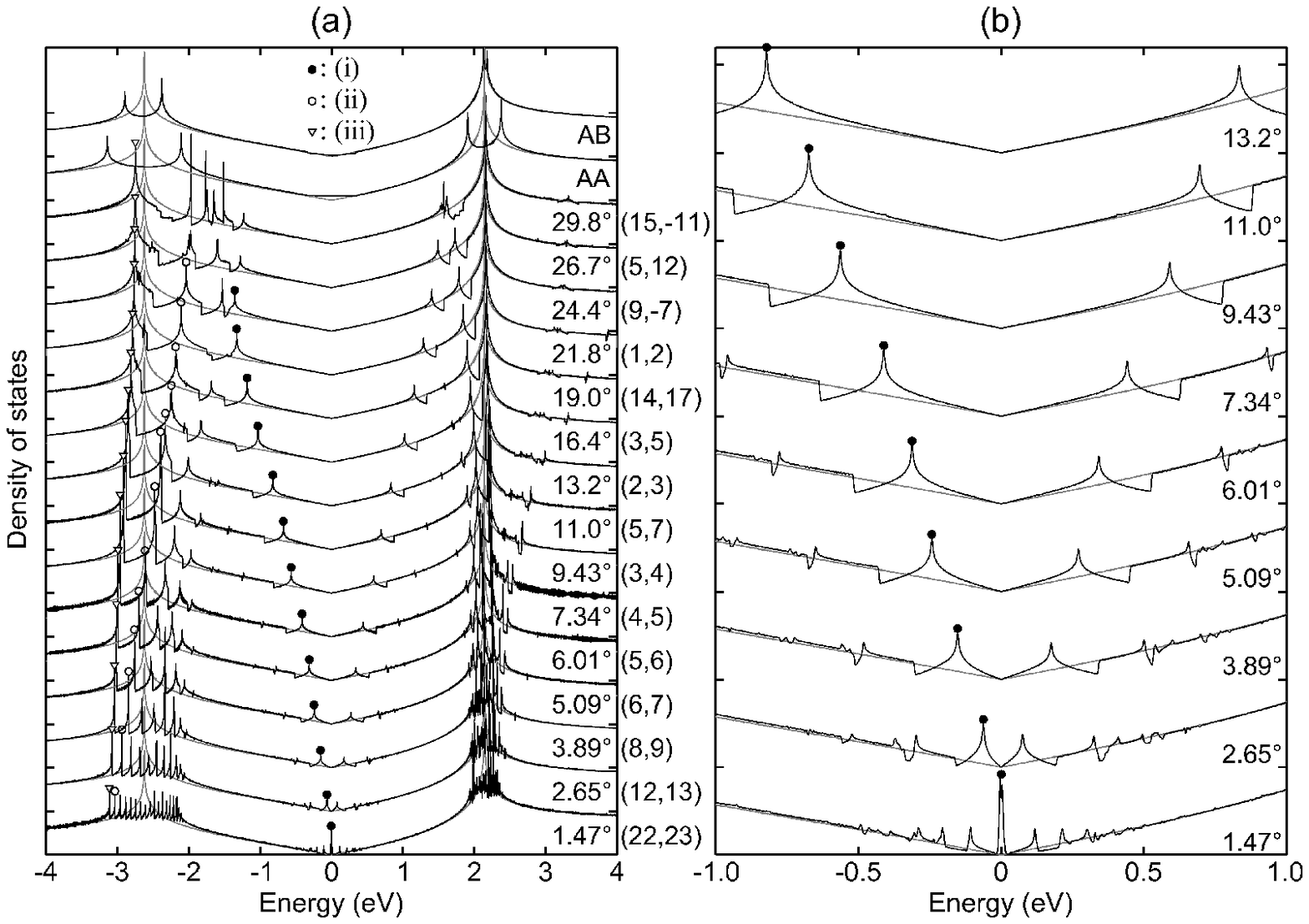}
\end{center}
\caption{DOS of TBGs 
with various rotation angles $0^\circ < \theta < 30^\circ$
in (a) wide and (b) narrow ranges of the energy.
The DOS of uncoupled bilayer graphene (i.e., twice of the monolayer's
 DOS) is shown as light gray lines.
Peaks marked with symbols
correspond to three different types of van Hove singularity (see text).
}
\label{fig_DOS_angle_dependence}
\end{figure*}

The band dispersion of TBG
can be intuitively understood in terms of
coupled four Dirac cones centered at the valleys
$K^{(l)}$ and $K'^{(l)}$ ($l = 1, 2$).
Figure \ref{fig_schematic_valley_coupling} shows
the schematic band structures of TBG
in the extended Brillouin zone
along the line $K'^{(2)}$-$K'^{(1)}$-$K^{(2)}$-$K^{(1)}$,
where Dashed and solid dispersion 
represent the energy bands of the layer 1 and 2, respectively.
The interlayer coupling gives rise to band anticrossing 
at the intersection,
and the resultant energy band is characterized by 
saddle points accompanied by the van Hove singularity
in the density of states. The saddle points are classified into
(i), (ii) and (iii) as shown in Fig.\ \ref{fig_schematic_valley_coupling},
where (i) and (ii) result from the band intersection 
of different layers, while (iii) originates 
from the original monolayer's band structure.
In the limit of $\theta \to 0$,
$K^{(1)}$ and $K^{(2)}$ (equivalently, $K'^{(1)}$ and $K'^{(2)}$)
get closer, so that (i) approaches the Dirac points,
while (ii) goes to the midpoint between $K$ and $K'$
in the high energy region.
(iii) remains at the constant energy.

Figures \ref{fig_band_structure_and_DOS_389}(a)
and \ref{fig_band_structure_and_DOS_389}(b)
show the band structure and
the density of state (DOS), respectively,
of TBG with $\theta=3.89^\circ$ [$(m,n)=(8,9)$]
actually calculated by the tight-binding model
Eq.\ (\ref{eq_Hamiltonian_TBG}).
The band structure can be viewed as monolayer's Dirac
cone folded into the superlattice Brillouin zone
with some band anticrossing at the zone corner.
\cite{latil2007massless,lopes2007graphene,shallcross2008quantum,hass2008multilayer,sprinkle2009first,morell2010flat,PhysRevB.85.195458}
The lowest band exhibits a linear dispersion
near $\bar{K}$ and $\bar{K}'$,
while we see a large splitting near the $\bar{M}$ point
in energy from $\pm$0.2eV to $\pm$0.4eV, respectively,
corresponding to the band anticrossing (i)
in Fig.\ \ref{fig_schematic_valley_coupling}.
\cite{li2009observation,PhysRevB.85.195458,ohta2012evidence}
The DOS has a sharp peak around $\pm0.2\,\rm eV$,
which is associated with the saddle point near the $\bar{M}$ point.

Each energy band can be classified
into either of those originating from
monolayer's $K$ region (i.e.,  $K^{(1)}$ and $K^{(2)}$)
or those from $K'$ region (i.e.,  $K'^{(1)}$ and $K'^{(2)}$),
because two valleys are hardly mixed by the interlayer interaction
in this small rotation angle.
Here $K$ and $K'$ should not be confused with
$\bar{K}$ and $\bar{K'}$ for the folded Brillouin zone.
The monolayer's band near $K$ and that near $K'$ are independently
folded into the same Brillouin zone
without mixing with each other.
Indeed, the lowest band in Fig.\ \ref{fig_band_structure_and_DOS_389}(a) 
is composed of nearly degenerate branches,
where dashed (pink) and solid (blue) lines
are the bands from $K$ and $K'$, respectively.
These two bands are degenerate 
along $\bar{K}-\bar{\Gamma}$ and $\bar{M}-\bar{K'}$,
reflecting the $C_2$ symmetry in the real-space lattice structure.
\cite{PhysRevB.85.195458}.

Figure \ref{fig_band_structure_angle_dependence}
shows the energy bands of TBGs with
different rotation angles from $\theta=9.43^\circ$ down to $1.47^\circ$.
The structures are similar to each other
while the overall energy scale shrinks as the rotation angle 
decreases, roughly in proportion to the size of the folded Brillouin zone.
The width of splitting at the $\bar{M}$ point
is about $0.2$eV in every case,
which is of the order of the interlayer coupling $V^0_{pp\sigma}$.
In small rotation angles less than $2^\circ$,
the energy scale of the folded Dirac cone
becomes comparable to the band splitting
so that the band velocity near the Dirac cone is significantly reduced
from the monolayer's.
\cite{lopes2007graphene,hass2008multilayer,ni2008reduction,morell2010flat,shallcross2010electronic,trambly2010localization,bistritzer2011moirepnas,PhysRevB.86.155449}
In Fig.\ \ref{fig_band_structure_angle_dependence},
we also plot the band energies calculated by the effective
continuum model Eq.\ (\ref{eq_eff_model})
as dashed (red) lines, to be compared with 
solid (black) curves obtained by the original tight-binding model.
We see the low-energy band structure agrees quite well,
except that the effective model fails to reproduce
a small electron hole asymmetry in the original model,
since it assumes the symmetric Dirac cone for the intralayer Hamiltonian.

Figure \ref{fig_DOS_angle_dependence}
shows the DOS of TBGs with various rotation angles
$0^\circ < \theta < 30^\circ$
with (a) wide and (b) narrow energy ranges.
To each curve, we append the DOS of
uncoupled bilayer graphene (i.e., twice of monolayer's)
as a light gray line.
We observe a number of characteristic peaks
associated with van Hove singularities (i), (ii) and (iii)
argued in Fig.\ \ref{fig_schematic_valley_coupling}.
As the rotation angle increases,
the peaks (i) move away from the Dirac points,
and the peaks (ii) move towards the Dirac points,
while the peaks (iii) stay at almost constant energy. 
In small rotation angles, we see a number of additional peaks
since the interlayer coupling
of the higher order in the Moir\'{e} wave number
becomes significant.

\subsection{Optical absorption}
\label{subsection:optical_selection_rule}

\begin{figure}[ht]
\begin{center}
\leavevmode\includegraphics[width=0.9\hsize]{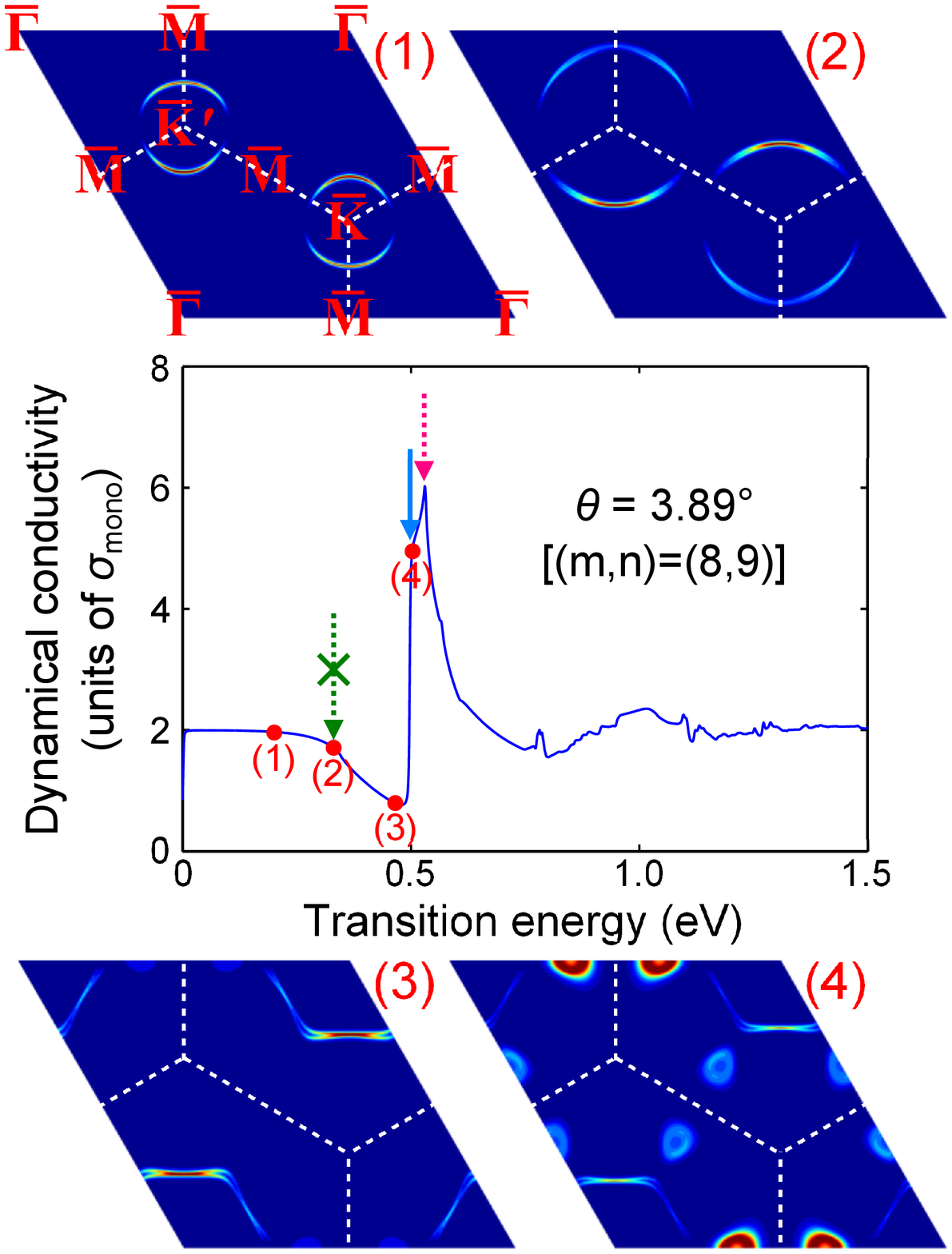}
\end{center}
\caption{(Middle panel) 
Dynamical conductivity of TBG with $\theta=3.89^\circ$
as a function of the transition energy.
Arrows indicate the excitation energies
of the transitions shown in Fig.\ \ref{fig_band_structure_and_DOS_389}(a).
(Top and bottom panels)
Spectral weight maps in the superlattice Brillouin zone
at several transition energies.
}
\label{fig_optical_conductivity_389}
\end{figure}

\begin{figure*}[ht]
\begin{center}
%\leavevmode\includegraphics[width=1.\hsize]{fig_optical_conductivity_angle_dependence.eps}
\leavevmode\includegraphics[width=1.\hsize]{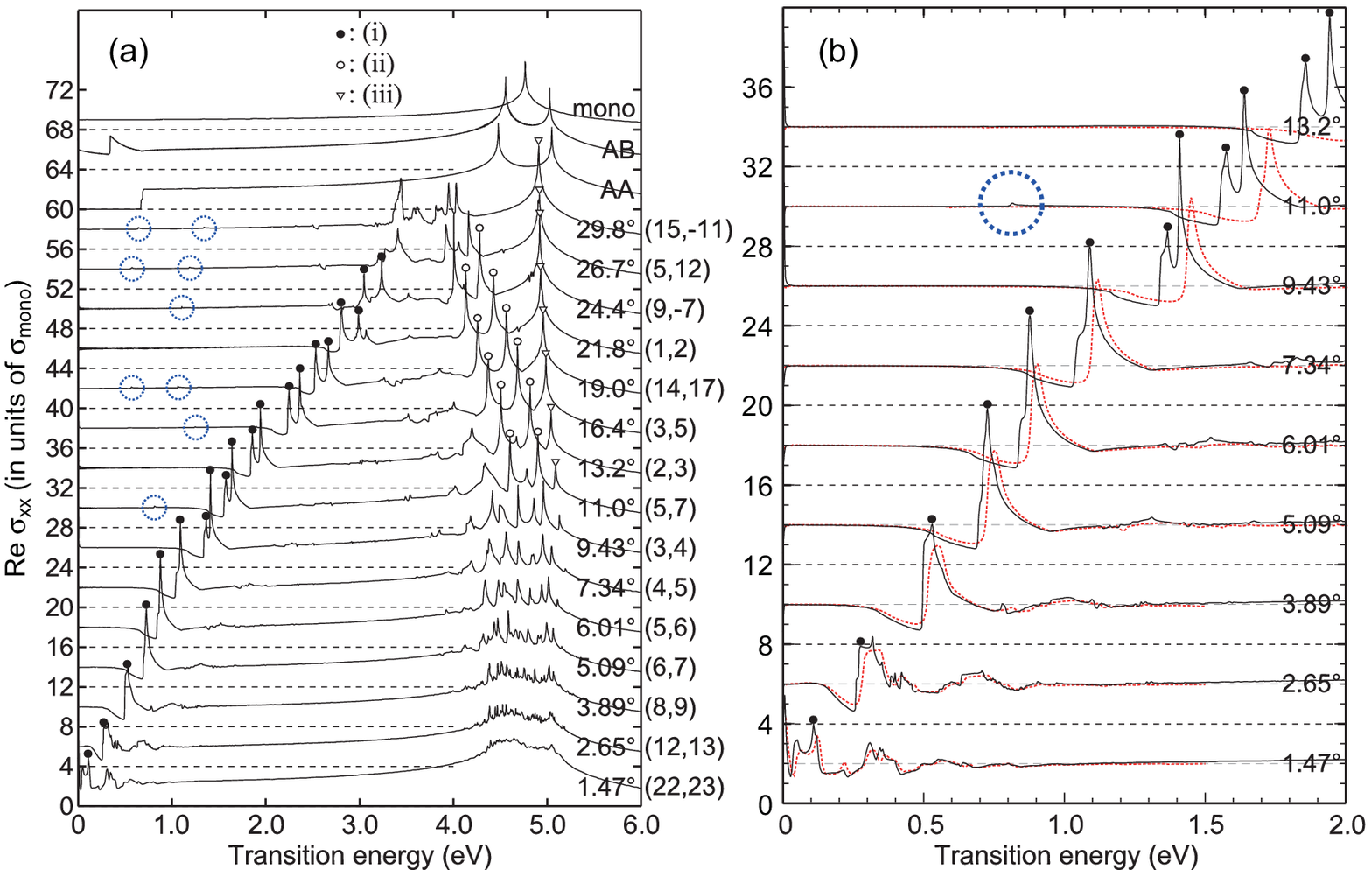}
\end{center}
\caption{
Dynamical conductivities of TBGs with various rotation angles
in (a) wide and (b) narrow frequency ranges,
calculated by the tight-binding model (solid black lines)
and the effective continuum model (dashed red lines, only for (b)).
Peaks marked with symbols
represent the excitations associated with 
the van Hove singularity in Fig.\ \ref{fig_DOS_angle_dependence}.
Dashed (blue) circles indicate the tiny peaks
which appears only when the actual lattice period $L$ 
is larger than the Moir\'{e} period $L_{\rm M}$ (see text).
}
\label{fig_optical_conductivity_angle_dependence}
\end{figure*}

In Fig.\ \ref{fig_optical_conductivity_389},
we plot the dynamical conductivity of TBG with $\theta=3.89^\circ$
calculated by the tight-binding model.
The conductivity is plotted in units of 
\begin{eqnarray}
\sigma_{\rm mono} = \frac{g_v g_s}{16} \frac{e^2}{\hbar}
\end{eqnarray}
which is the universal dynamical conductivity of monolayer graphene
at linear band regime, 
where $g_s=2$ and $g_v=2$ are the spin and valley ($K$, $K'$)
degeneracy, respectively. 
\cite{ando2002dynamical,gusynin2006transport,gusynin2006unusual,gusynin2007anomalous,nair2008fine}
The spectrum is characterized by a peak around $0.5\,\rm eV$,
and considerable reduction right below the peak energy.
Otherwise the conductivity is close to $2\sigma_{\rm mono}$.
The top and bottom panels numbered from (1) to (4)
show spectral weight maps at specific photon energies,
which highlight the wave vectors that contribute to the optical transition. 
A sudden rise in the conductivity between (3) and (4)
is due to bright spots near $\bar{M}$ point in the weight map.
This actually corresponds to the transition
from the saddle point in the lowest valence band to 
the second conduction band,
which is marked with a solid (blue) arrow
in the band diagram in Fig.\ \ref{fig_band_structure_and_DOS_389}(a).
The similar transition 
%from the second valence band to the first conduction band,
indicated by dashed (pink) arrow
occurs at slightly higher energy due to the electron-hole asymmetry.
Below the peak energy, the dynamical conductivity is significantly reduced 
and becomes lower than $2\sigma_{\rm mono}$,
because the number of available states 
largely decreases due to the band anticrossing.

It should be noted that the transition 
does not occur between the saddle points
of the lowest conduction and valence bands 
[dotted (green) arrow in Fig.\ \ref{fig_band_structure_and_DOS_389}(a)],
because it is optically forbidden
in this particular system.
This can be clearly explained by the 
effective continuum model as following.
The lowest electron and hole states at $\bar{M}$ point,
which are connected by dotted arrow in Fig.\ \ref{fig_band_structure_and_DOS_389},
are an electron-hole pair 
related by the symmetry of Eq.\ (\ref{eq_e-h_symmetry}),
and thus these wavefunctions are written as
$\psi$ and $\hat{\Sigma} \psi^*$.
The matrix element of $v_x$ between the two states
is obviously zero,
because 
\begin{eqnarray}
&& \langle \hat{\Sigma} \psi^* | v_x | \psi \rangle
=
\begin{pmatrix}
\psi^*_4 \\
\psi^*_3 \\
-\psi^*_2\\
-\psi^*_1
\end{pmatrix}^\dagger
\begin{pmatrix}
v \sigma_x & 0\\
0 & v \sigma_x 
\end{pmatrix}
\begin{pmatrix}
\psi_1 \\
\psi_2 \\
\psi_3\\
\psi_4
\end{pmatrix}
\nonumber\\
&&
= v(\psi_4 \psi_2 + \psi_3 \psi_1 - \psi_2 \psi_4 -\psi_1 \psi_3)
=0,
\end{eqnarray}
and thus the transition is optically inactive.
This symmetry does not limit
the optical selection rule at other points than $\bar{M}$
[the origin of the wave number in the 
effective Hamiltonian Eq.\ (\ref{eq_eff_model})],
since $\psi$ and $\hat{\Sigma} \psi^*$ generally reside
at different Bloch wave vectors 
and are not connected by the optical transition.

In Fig.\ \ref{fig_optical_conductivity_angle_dependence}(a),
we plot the optical absorption spectra
at various rotation angles in a wide frequency range.
The spectrum exhibits characteristic conductivity peaks
ranging from terahertz to ultraviolet frequencies.
The peaks are again classified into three groups
similarly to the density of states.
When the rotation angle increases from $0^\circ$ to $30^\circ$,
the peak (i) [(ii)] moves to higher (lower) energies,
while the peak (iii) remains unchanged.
%In large angles $\theta > 10^\circ$,
%the peak (i) splits due to the electron-hole asymmetry.
Figure \ref{fig_optical_conductivity_angle_dependence}(b)
shows magnified plots of the low-frequency range
for several small angles.
There the spectrum is characterized by a single peak belonging to group (i)
similarly to Fig.\ \ref{fig_optical_conductivity_389},
and its transition energy monotonically 
shifts with the rotation angle.
In $\theta = 1.47^\circ$, the spectrum exhibits a complicated structure
in accordance with the strong band deformation 
observed in Fig.\ \ref{fig_band_structure_angle_dependence}.
In Fig.\ \ref{fig_optical_conductivity_angle_dependence}(b), 
we also present the spectrum of 
the effective continuum model as dashed (red) curves,
to be compared with the original tight-binding calculation.
The results agree quite well except for the
peak splitting 
due to the electron-hole asymmetry,
which is pronounced in $\theta \gsim 10^\circ$.

As argued in Sec.\ \ref{subsection:Atomic structure and Brillouin zone},
the rigorous superlattice period $L$
in Eq.\ (\ref{eq_superlattice_period}) discontinuously changes
depending on the commensurability of the lattice periods.
%which is a continous function of 
%the rotation angle $\theta$.
Nevertheless, the DOS in Fig.\ \ref{fig_DOS_angle_dependence}
and the optical absorption spectrum
in Fig.\ \ref{fig_optical_conductivity_angle_dependence}
show that the peak structure almost continuously evolves 
with the rotation angle,
suggesting that the exact lattice commensurability is not quite important
for the physical property.
%This is also consistent with the fact that 
%the effective continuum model works well.
Although we cannot rigorously handle
incommensurate TBGs due to infinite unit cell size,
its optical spectrum should be approximated
by interpolating those of commensurate TBGs
with similar rotation angles.
The only property in which the actual lattice period $L$ matters
is found as tiny peaks in the conductivity
indicated as dashed (blue) circles in 
Fig.\ \ref{fig_optical_conductivity_angle_dependence},
which exist only when $|m-n|>1$.
They are related to the transition at the corner of the 
exact superlattice Brillouin zone.

%In addition to the rotation angle, also the relative translation $t$
%between the layers defines the stacking geometry of
%two graphene layers.
%Up to this point, we have considered the structures with
%a specific $t$, which gives the SE-odd atomic configuration
%($r = -1$, $S_1^{(l)}$ at the origin).
%However, we can make TBGs with different atomic structures
%by sliding the second layer with respect to the first one
%with an arbitrary translation vector.

\subsection{Effect of lattice displacement}

\begin{figure}
\begin{center}
\leavevmode\includegraphics[width=0.8\hsize]{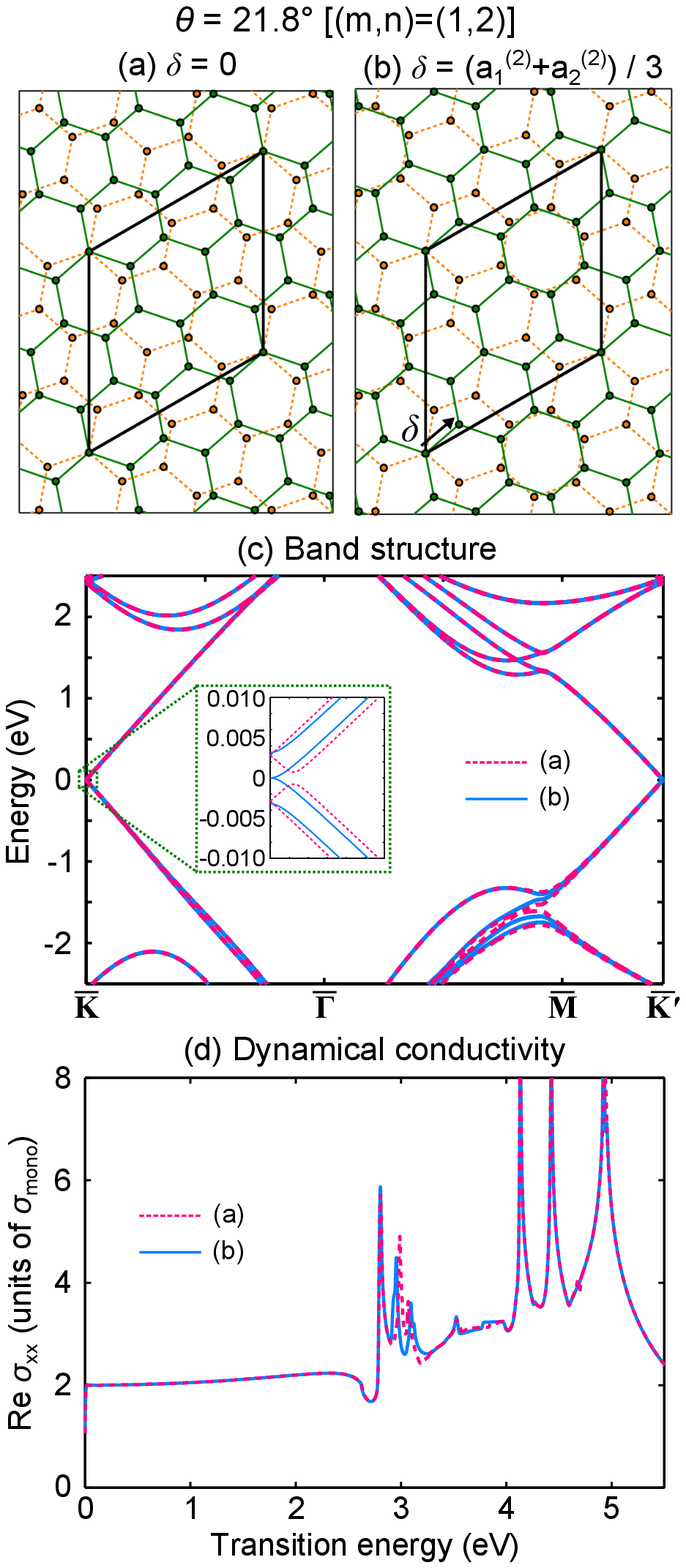}
\end{center}
\caption{Atomic structures of TBGs of $\theta=21.8^\circ$
with different translation vectors,
(a) $\GVec{\delta}=0$ and (b) $(\Vec{a}^{(2)}_1+\Vec{a}^{(2)}_2)/3$.
(c) Band structures and (d) dynamical conductivities of the two
distinct TBGs.}
\label{fig_band_structure_and_optical_conductivity_translation_dependence_218}
\end{figure}

The electronic structure depends on the lattice displacement 
$\GVec{\delta}$,
particularly when the superlattice period $L$ is comparable to the 
atomic scale.
This effect is maximum at $\theta = 0$, where
the lattice constant $L$ coincides with $a$,
and the AA stacking is transformed to the AB stacking by 
a translation $\GVec{\delta}$.
The TBG with $(m,n)=(1,2)$ ($\theta = 21.8^\circ$)
has the next smallest primitive unit cell of $L = \sqrt{7}a$.
Here we focus on two distinct TBGs of $(m,n)=(1,2)$
with $\GVec{\delta}=0$ and $(\Vec{a}^{(2)}_1+\Vec{a}^{(2)}_2)/3$,
shown in
Figs.\ \ref{fig_band_structure_and_optical_conductivity_translation_dependence_218}(a)
and (b), respectively,
to investigate the effect of the relative translation on
the optical absorption.
These two structures can be also described by
two different rotation angles $\theta$ and $60^\circ - \theta$.
Figures \ref{fig_band_structure_and_optical_conductivity_translation_dependence_218}(c) and (d) show the electronic structure
and the dynamical conductivity, respectively, 
of the two different TBGs.
We observe that the energy bands are almost
equivalent, while there is a tiny difference
in the structure near $\bar{M}$ point
of which the energy scale is a few tens of meV.
In the dynamical conductivity, this is reflected in a small difference
in the peak structure near the energy of $3$ eV.
The inset in
Fig.\
\ref{fig_band_structure_and_optical_conductivity_translation_dependence_218}(c) magnifies the band structure in the vicinity the Dirac points. 
We observe a difference of the energy scale about a few meV,
where one takes the form like AA-bilayer and the other 
like AB-bilayer, \cite{shallcross2008quantum,mele2010commensuration,mele2012interlayer}
while it gives no noticeable difference in the
dynamical conductivity in
Fig.\ \ref{fig_band_structure_and_optical_conductivity_translation_dependence_218}(d).
The dependence on the lattice displacement
becomes even smaller as the superlattice period $L$
becomes larger, and is completely absent in any incommensurate angles.
Thus, concerning the energy range of interest,
we conclude that the optical spectrum of TBG does not depends much on 
the lattice displacement, except for $\theta =0$.

\section{CONCLUSION}
\label{section:conclusion}

We theoretically investigated 
the optical absorption properties of TBGs
with various stacking geometries
using the tight-binding model and the effective continuum model.
We showed that the spectrum is 
characterized by series of absorption peaks 
associated with the van Hove singularities 
in the band structure,
and the peak energies systematically shift
in changing the rotation angle.
The optical spectrum almost continuously evolves 
in changing the rotation angle 
regardless of the rigorous commensurability between two layers,
suggesting that the optical absorption measurement
provides a convenient way to identify the rotation angle of TBG.
We developed the effective continuum model
based on the tight-binding model used here,
and demonstrated that it well reproduces 
the low-energy band structure and the dynamical conductivity 
of the tight-binding model for $\theta < 10^\circ$,
and it also explains the optical selection rule analytically
in terms of the symmetry of the effective Hamiltonian.

\section*{ACKNOWLEDGMENTS}

This project has been funded by 
Grant-in-Aid for Research Activity Start-up No. 23840004 (PM),
Grant-in-Aid for Scientific Research No. 24740193 (MK)
from Japan Society for the Promotion of Science (JSPS),
and by JST-EPSRC Japan-UK
Cooperative Programme Grant No. EP/H025804/1.
P.\ M.\ acknowledges 
the Supercomputer Center, Institute for Solid State Physics,
University of Tokyo for the use of the facilities (Project No. ID: H23-D-0009).

%\subsection{Appendix: Derivation of effective interlayer coupling}
\appendix
\section{Derivation of effective interlayer coupling}
\label{sec_app1}

Here we derive the effective interlayer matrix element
Eq.\ (\ref{eq_effective_interlayer})
for TBG with small rotation angles.
The matrix element of the Hamiltonian 
Eq.\ (\ref{eq_Hamiltonian_TBG})
between the single layer bases 
on the different layers
is explicitly written as
\begin{eqnarray}
\langle \Vec{k}',X'_2| H |\Vec{k},X_1\rangle
&&
\nonumber\\
&&\hspace{-30mm}
=
\frac{1}{N}\sum_{\Vec{R}_{X'_2},\Vec{R}_{X_1}}
- t(\Vec{R}_{X'_2}-\Vec{R}_{X_1})
\exp[
-i\Vec{k}'\cdot\Vec{R}_{X'_2}+i\Vec{k}\cdot\Vec{R}_{X_1}
]
\nonumber\\
&&\hspace{-30mm}
=
\frac{1}{N}\sum_{\Vec{R}_{X'_2},\Vec{R}_{X_1}}
- t(\Vec{R}_{X'_2}-\Vec{R}_{X_1})
\nonumber\\
&&\hspace{-25mm}
\times 
\exp\left[
-i\bar{\Vec{k}}\cdot(\Vec{R}_{X'_2}-\Vec{R}_{X_1})
\right]
\exp\left[
-i\Delta\Vec{k}\cdot\frac{\Vec{R}_{X'_2}+\Vec{R}_{X_1}}{2}
\right].
\nonumber\\
\label{eq_app1}
\end{eqnarray}
where
\begin{eqnarray}
 \bar{\Vec{k}} = \frac{\Vec{k}+\Vec{k}'}{2},
\quad
\Delta\Vec{k} = \Vec{k}'-\Vec{k}.
\end{eqnarray}

Since the Moire lattice constant $L_{\rm M}$ is much larger
than $a$ in small rotation angles,
we only need to 
consider $\Vec{k}$ and $ \Vec{k}'$ which are close to each other,
or $|\Delta\Vec{k}| \ll 2\pi/a$.
Then the last exponential term in Eq.\ (\ref{eq_app1}) 
slowly varies in the lattice position,
while other terms change in the atomic length scale.
To separate out the long wave component,
we introduce a smoothing function $g(\Vec{r})$
which satisfies the following conditions: \cite{ando2005theory}
 $g(\Vec{r})$ varies in $\Vec{r}$ in an intermediate length scale $l_g$,
which is much larger than the lattice constant
$a$, but much smaller than the Moir\'{e} superlattice period $L_{\rm M}$.
$g(\Vec{r})$ is a peak centered at $\Vec{r}=0$,
and rapidly decays in $\Vec{r} \gsim l_g$. The area is normalized as
\begin{eqnarray}
 \int g(\Vec{r}) d\Vec{r} = \Omega_{\rm M}, 
 \label{eq_g_int}
\end{eqnarray}
where $\Omega_{\rm M}$ is the 
Moir\'{e} superlattice unit cell,
and the integral is taken over the whole system area 
unless otherwise stated. 
Almost equivalently, we have
\begin{eqnarray}
\sum_{\Vec{R}_{X}} g(\Vec{r}-\Vec{R}_{X}) = 
 \frac{\Omega_{\rm M}}{\Omega_0},
 \label{eq_g_sum}
\end{eqnarray}
where $X$ is either of $A_1,B_1,A_2$ or $B_2$,
and $\Omega_0 = |\Vec{a}_1\times\Vec{a}_2|$
is the area of monolayer's unit cell.

Using Eq.\ (\ref{eq_g_sum}), the matrix element Eq.\ (\ref{eq_app1})
is written as
\begin{eqnarray}
\langle \Vec{k}',X'_2| H |\Vec{k},X_1\rangle
&&
\nonumber\\
&&\hspace{-29mm}
=
\frac{1}{N}
\sum_{\Vec{R}_{X'_2},\Vec{R}_{X_1}}
\left[
\frac{1}{\Omega_{\rm M}} \int g(\Vec{r}-\Vec{R}_{X_1}) d\Vec{r}
\right]
\left[
- t(\Vec{R}_{X'_2}-\Vec{R}_{X_1})
\right]
\nonumber\\
&&\hspace{-25mm}
\times 
\exp\left[
-i\bar{\Vec{k}}
\cdot(\Vec{R}_{X'_2}-\Vec{R}_{X_1})
\right]
\exp\left[
-i\Delta\Vec{k}\cdot\frac{\Vec{R}_{X'_2}+\Vec{R}_{X_1}}{2}
\right].
\nonumber\\
\end{eqnarray}
Now the argument $(\Vec{R}_{X'_2}+\Vec{R}_{X_1})/2$ 
in the last can be replaced with $\Vec{R}_{X_1}$,
because  the last exponential term 
varies slowly with the length scale of $L_{\rm M}$,
and also the hopping integral $t(\Vec{R}_{X'_2}-\Vec{R}_{X_1})$
occurs only in the atomic scale distance.
This is further replaced with $\Vec{r}$,
since the smoothing factor $g(\Vec{r}-\Vec{R}_{X_1})$
works as the delta function 
for the slowly varying function with the scale $L_{\rm M}$.
By including this, we obtain
\begin{eqnarray}
\langle \Vec{k}',X'_2| H |\Vec{k},X_1\rangle
&&
\nonumber\\
&&\hspace{-30mm}
\approx
\frac{1}{N \Omega_{\rm M}} 
\int d\Vec{r} \,
e^{-i\Delta\Vec{k}\cdot\Vec{r}}
\sum_{\Vec{R}_{X_1}}
 g(\Vec{r}-\Vec{R}_{X_1}) 
\nonumber\\
&&\hspace{-25mm}
\times 
\sum_{\Vec{R}_{X'_2}}
\left[
- t(\Vec{R}_{X'_2}-\Vec{R}_{X_1})
\right]
\exp\left[
-i\bar{\Vec{k}}\cdot(\Vec{R}_{X'_2}-\Vec{R}_{X_1})
\right]
\nonumber
\\
&&\hspace{-30mm}
=\frac{1}{\Omega_{\rm M}} 
\int_{\Omega_{\rm M}} d\Vec{r}\,
e^{-i\Delta\Vec{k}\cdot\Vec{r}}
U_{X'_2X_1}
\left[\bar{\Vec{k}},\GVec{\delta}(\Vec{r})
\right],
\end{eqnarray}
where we used Eq.\ (\ref{eq_g_sum})
and $\int d\Vec{r}/(N \Omega_0) 
= \int_{\Omega_{\rm M}} d\Vec{r}/\Omega_{\rm M}$.
The last equation is the final result of
Eq.\ (\ref{eq_effective_interlayer}).

\bibliography{Qiqqa2BibTexExport}

\end{document}